\documentclass[11pt,a4paper]{article}
\usepackage{jheppub}
\usepackage{amsmath}
\usepackage{amssymb}
\usepackage{braket}
\usepackage[normalem]{ulem}
\usepackage{bm}

\newcommand{\cse}{\mathcal{C} \subseteq \mathcal{E}}

\title{Geometric Constraints from Subregion Duality Beyond the Classical Regime}

\author{Chris Akers,}
\author{Jason Koeller,}
\author{Stefan Leichenauer,}
\author{and Adam Levine}

\affiliation{Center for Theoretical Physics and Department of Physics,\\
University of California, Berkeley, CA 94720, U.S.A. and}
\affiliation{Lawrence Berkeley National Laboratory, Berkeley, CA 94720, U.S.A.}

\abstract{
Subregion duality in AdS/CFT implies certain constraints on the geometry: entanglement wedges must contain causal wedges, and nested boundary regions must have nested entanglement wedges. We elucidate the logical connections between these statements and the Quantum Focussing Conjecture, Quantum Null Energy Condition, Boundary Causality Condition, and Averaged Null Energy Condition. Our analysis does not rely on the classical limit of bulk physics, but instead works to all orders in $G\hbar \sim 1/N$. This constitutes a nontrivial check on the consistency of subregion duality, entanglement wedge reconstruction, and holographic entanglement entropy beyond the classical regime.
}

\begin{document}

\maketitle


\section{Introduction}

AdS/CFT implies constraints on quantum gravity from properties of quantum field theory. For example, field theory causality requires that null geodesics through the bulk are delayed relative to those on the boundary. Such constraints on the bulk geometry can often be understood as coming from energy conditions on the bulk fields. In this case, bulk null geodesics will always be delayed as long as there is no negative null energy flux \cite{Gao:2000ga}.

In this paper, we examine two constraints on the bulk geometry that are required by the consistency of the AdS/CFT duality. The starting point is the idea of subregion duality, which is the idea that the state of the boundary field theory reduced to a subregion $A$ is itself dual to a subregion of the bulk. The relevant bulk region is called the entanglement wedge, $\mathcal{E}(A)$, and consists of all points spacelike related to the extremal surface anchored on \(\partial A\), on the side towards \(A\) \cite{Czech:2012bh,Headrick:2014cta}. The validity of subregion duality was argued \cite{Dong:2016eik,Harlow:2016vwg} to follow from the Ryu-Takayanagi-FLM formula \cite{Ryu:2006ef,Ryu:2006bv,Hubeny:2007xt,Faulkner:2013ana,Lewkowycz:2013nqa,Dong:2016hjy}, and the consistency of subregion duality immediately implies two constraints on the bulk geometry.

The first constraint, which we call Entanglement Wedge Nesting (EWN), is that if a region $A$ is contained in a region $B$ on the boundary (or more generally, if the domain of dependence of $A$ is contained in the domain of dependence of $B$), then $\mathcal{E}(A)$ must be contained in $\mathcal{E}(B)$. This condition was previously discussed in \cite{Czech:2012bh,Wall:2012uf}. 

The second constraint is that the set of bulk points $I^-(D(A)) \cap I^+(D(A))$, called  the causal wedge $\mathcal{C}(A)$, is completely contained in the entanglement wedge $\mathcal{E}(A)$. We call this $\mathcal{C} \subseteq \mathcal{E}$. See \cite{Czech:2012bh,Wall:2012uf,Engelhardt:2014gca,Hubeny:2012wa,Headrick:2014cta} for previous discussion of \(\cse\).

We refer to the delay of null geodesics passing through the bulk relative to their boundary counterparts \cite{Gao:2000ga} as the Boundary Causality Condition (BCC), as in \cite{Engelhardt:2016aoo}. These three conditions, and their connections to various bulk and boundary inequalities relating entropy and energy, are the primary focus of this paper.

In section \ref{sec:glossary} we will spell out the definitions of EWN and $\mathcal{C} \subseteq \mathcal{E}$ in more detail, as well as describe their relations with subregion duality. Roughly speaking, EWN encodes the fact that subregion duality should respect inclusion of boundary regions. $\mathcal{C} \subseteq \mathcal{E}$ is the statement that the bulk region dual to a given boundary region should at least contain all those bulk points from which messages can be both received from and sent to the boundary region. 

Even though EWN, $\mathcal{C} \subseteq \mathcal{E}$, and the BCC are all required for consistency of AdS/CFT, part of our goal is to investigate their relationships to each other as bulk statements independent of a boundary dual. As such, we will demonstrate that EWN implies $\mathcal{C} \subseteq \mathcal{E}$, and $\mathcal{C} \subseteq \mathcal{E}$ implies the BCC. Thus EWN is in a sense the strongest statement of the three.

Though this marks the first time that the logical relationships between EWN, $\mathcal{C} \subseteq \mathcal{E}$, and the BCC have been been independently investigated, all three of these conditions are known in the literature and have been proven from more fundamental assumptions in the bulk \cite{Wall:2012uf,Hubeny:2012wa}. In the classical limit, a common assumption about the bulk physics is the Null Energy Condition (NEC).\footnote{See \cite{Headrick:2014cta} for a related classical analysis of bulk constraints from causality, including \(\cse\).} However, the NEC is known to be violated in quantum field theory. Recently, the Quantum Focussing Conjecture (QFC), which ties together geometry and entropy, was put forward as the ultimate quasi-local ``energy condition'' for the bulk, replacing the NEC away from the classical limit \cite{Bousso:2015mna}.

The QFC is currently the strongest reasonable quasi-local assumption that one can make about the bulk dynamics, and indeed we will show below that it can be used to prove EWN. Additionally, there are other, weaker, restrictions on the bulk dynamics which follow from the QFC. The Generalized Second Law (GSL) of horizon thermodynamics is a consequence of the QFC. In \cite{Engelhardt:2014gca}, it was shown that the GSL implies what we have called $\mathcal{C} \subseteq \mathcal{E}$. Thus the QFC, the GSL, EWN, and $\mathcal{C} \subseteq \mathcal{E}$ form a square of implications. The QFC is the strongest of the four, implying the three others, while the $\mathcal{C} \subseteq \mathcal{E}$ is the weakest. This pattern continues in a way summarized by Figure \ref{fig:CHART}, which we will now explain. 

\begin{figure}\label{fig:CHART}
	\centering
	\includegraphics[width = \linewidth]{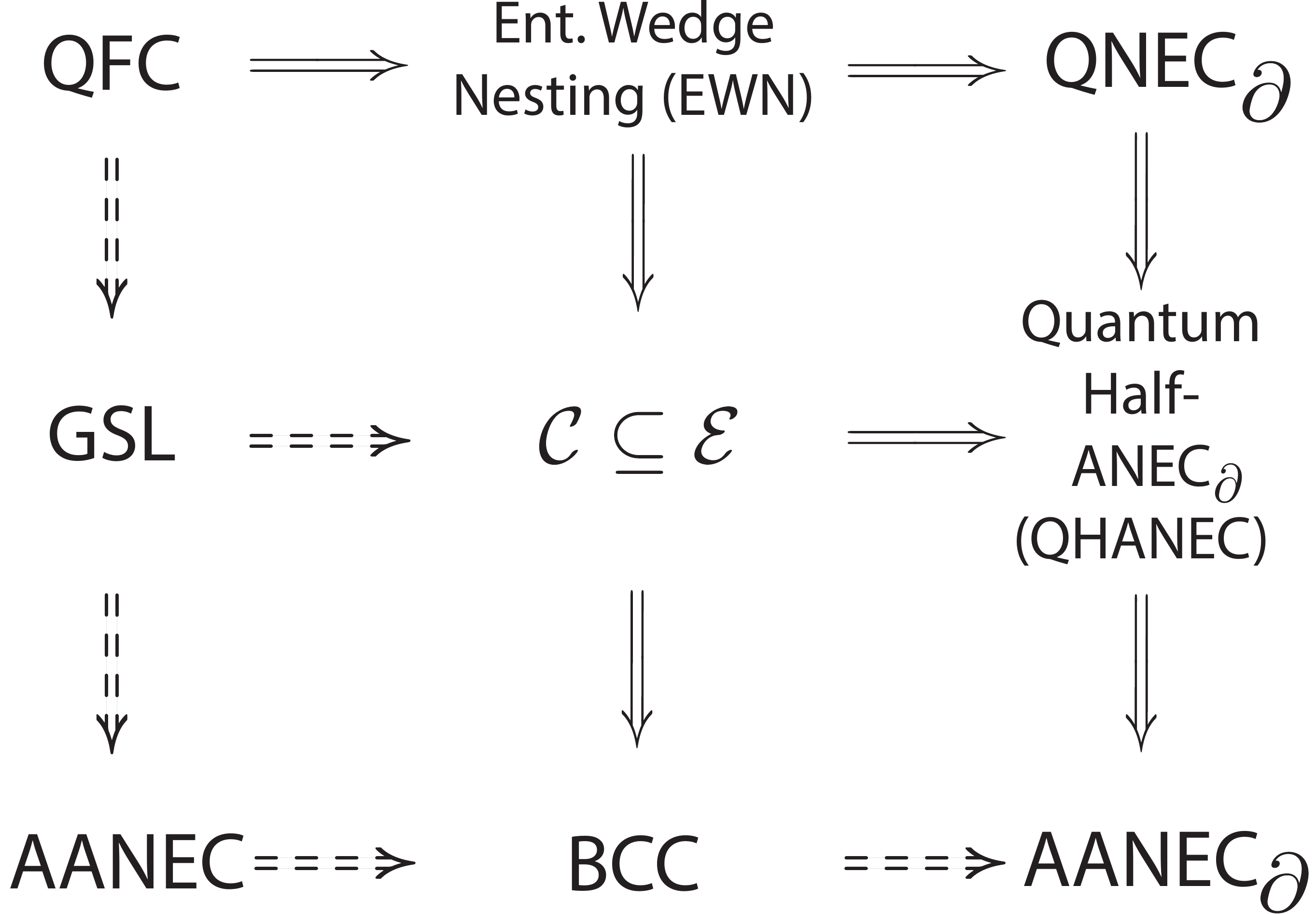}
	\caption{The logical relationships between the constraints discussed in this paper. The left column contains semi-classical quantum gravity statements in the bulk. The middle column is composed of constraints on bulk geometry. In the right column is quantum field theory constraints on the boundary CFT. All implications are true to all orders in $G\hbar \sim 1/N$. We have used dashed implication signs for those that were proven to all orders before this paper.}
	\label{fig:CHART}
\end{figure}

The QFC, the GSL, and the Achronal Averaged Null Energy Condition (AANEC) reside in the first column of Fig.~\ref{fig:CHART}. As we have explained, the QFC is the strongest of these three, while the AANEC is the weakest \cite{Wall:2009wi}. In the second column we have EWN, $\mathcal{C} \subseteq \mathcal{E}$, and the BCC. In addition to the relationships mentioned above, it was shown in \cite{Gao:2000ga} that the ANEC implies the BCC, which we extend to prove the BCC from the AANEC.

The third column of Figure \ref{fig:CHART} contains ``boundary'' versions of the first column: the Quantum Null Energy Condition (QNEC) \cite{Bousso:2015mna,Bousso:2015wca,Koeller:2015qmn}, the Quantum Half Averaged Null Energy Condition (QHANEC), and the boundary AANEC.\footnote{For simplicity we are assuming throughout that the boundary theory is formulated in Minkowski space. There would be additional subtleties with all three of these statements if the boundary were curved.} These are field theory statements which can be viewed as nongravitational limits of the corresponding inequalities in the first column. The QNEC is the strongest, implying the QHANEC, which in turn implies the AANEC. All three of these statements can be formulated in non-holographic theories, and all three are conjectured to be true generally. (The AANEC was recently proven in \cite{Faulkner:2016mzt} as a consequence of monotonicity of relative entropy and in \cite{Hartman:2016lgu} as a consequence of causality.)

In the case of a holographic theory, it was shown in \cite{Koeller:2015qmn} that EWN in the bulk implies the QNEC for the boundary theory to leading order in $G\hbar \sim 1/N$. We demonstrate that this relationship continues to hold with bulk quantum corrections. Moreover, in \cite{Kelly:2014mra} the BCC in the bulk was shown to imply the boundary AANEC. Here we will complete the pattern of implications by showing that $\mathcal{C} \subseteq \mathcal{E}$ implies the boundary QHANEC.

In the classical regime, the entanglement wedge is defined in terms of a codimension-2 surface with extremal area \cite{Hubeny:2007xt,Faulkner:2013ana,Headrick:2014cta,Dong:2016hjy}. It has been suggested that the correct quantum generalization should be defined in terms of the ``quantum extremal surface'': a Cauchy-splitting surface which extremizes the generalized entropy to one side \cite{Engelhardt:2014gca}. Indeed, we find that the logical structure of Fig.~\ref{fig:CHART} persists to all orders in perturbation theory in \(G\hbar \sim 1/N\) if and only if the entanglement wedge is defined in terms of the quantum extremal surface. This observation provides considerable evidence for prescription of \cite{Engelhardt:2014gca}.

The remainder of this paper is organized as follows. In Section \ref{sec:glossary} we will define all of the statements we set out to prove, as well as establish notation. Then in Sections \ref{sec:energyentropy} and \ref{sec:proofs} we will prove the logical structure encapsulated in Figure \ref{fig:CHART}. Several of these implications are already established in the literature, but for completeness we will briefly review the relevant arguments. We conclude with a discussion in Section \ref{sec:discussion}.

\section{Glossary}\label{sec:glossary}
\paragraph{Regime of Validity}
Quantum gravity is a tricky subject. We work in a semiclassical (large-\(N\)) regime, where the dynamical fields can be expanded perturbatively in \(G \hbar \sim 1/N\) about a classical background \cite{Wall:2011hj}.\footnote{The demensionless expansion parameter would be $G\hbar/\ell^{D-2}$, where $\ell$ is a typical length scale in whatever state we are considering. We will leave factors of $\ell$ implicit.} For example, the metric has the form
\begin{align}
	g_{ab} = g^{0}_{ab} + g^{1/2}_{ab} + g^{1}_{ab} + O((G\hbar)^{3/2})~,
\end{align}
where the superscripts denote powers of \(G \hbar\). In the semi-classical limit --- defined as \(G\hbar \to 0\) --- the validity of the various inequalities we consider will be dominated by their leading non-vanishing terms. We assume that the classical \(O((G\hbar)^0)\) part of the metric satisfies the null energy condition (NEC), without assuming anything about the quantum corrections. For more details on this type of expansion, see Wall \cite{Wall:2009wi}.

We primarily consider the case where the bulk theory can be approximated as Einstein gravity with minimally coupled matter fields. In the semiclassical regime, bulk loops will generate Planck-suppressed higher derivative corrections to the gravitational theory and the gravitational entropy.\footnote{Such corrections are also necessary for the generalized entropy to be finite. See Appendix A of \cite{Bousso:2015mna} for details and references. Other terms can be generated from, for example, stringy effects, but these will be suppressed by the string length $\ell_s$. For simplicity, we will not separately track the $\ell_s$ expansion. This should be valid as long as the string scale is not much different from the Planck scale.} We will comment on the effects of these corrections throughout.

We consider a boundary theory on flat space, possibly deformed by relevant operators. When appropriate, we will assume the null generic condition, which guarantees that every null geodesic encounters matter or gravitational radiation.

\subsection{Geometrical Constraints}\label{sec:gloss:geometrical}
There are a number of known properties of the AdS bulk causal structure and extremal surfaces. At the classical level (i.e. at leading order in \(G\hbar \sim 1/N\)), the NEC is the standard assumption made about the bulk which ensures that these properties are true \cite{Wall:2012uf}. However, some of these are so fundamental to subregion duality that it is sensible to demand them and to ask what constraints in the bulk might ensure that these properties hold, even under quantum corrections. Answering this question is one key focus of this paper. 

In this section, we review three necessary geometrical constraints. In addition to defining each of them and stating their logical relationships (see Figure \ref{fig:CHART}), we explain how each is critical for subregion duality.

\subsubsection*{Boundary Causality Condition (BCC)}
A standard notion of causality in asymptotically-AdS spacetimes is the condition that \emph{the bulk cannot be used for superluminal communication relative to the causal structure of the boundary}. More precisely, any causal bulk curve emanating from a boundary point $p$ and arriving back on the boundary must do so to the future of $p$ as determined by the boundary causal structure.

This condition, termed ``BCC'' in \cite{Engelhardt:2016aoo}, is known to follow from the averaged null curvature condition (ANCC) \cite{Gao:2000ga}. Engelhardt and Fischetti have derived an equivalent formulation in terms of an integral inequality for the metric perturbation in the context of linearized perturbations to the vacuum \cite{Engelhardt:2016aoo}.

Microcausality in the boundary theory requires that the BCC hold. If the BCC were violated, a bulk excitation could propagate between two spacelike-separated points on the boundary leading to nonvanishing commutators of local fields at those points. In Sec. \ref{sec:proofs} we will show that BCC is implied by $\mathcal{C} \subseteq \mathcal{E}$. Thus BCC is the weakest notion of causality in holography that we consider.

\subsubsection*{$\bm{\mathcal{C} \subseteq \mathcal{E}}$}
Consider the domain of dependence $D(A)$ of a boundary region $A$. Let us define the causal wedge of a boundary region $A$ to be $\mathcal{C}(A) \equiv I^-(D(A)) \cap I^+(D(A))$.\footnote{\(I^{\pm}(S)\) represent respectively the chronological future and past of the set \(S\). The causal wedge was originally defined in \cite{Headrick:2014cta} in terms of the causal future and past, \(J^{\pm}(S)\), but for our purposes the chronological future and past are more convenient.}

By the Ryu-Takayanagi-FLM formula, the entropy of the quantum state restricted to $A$ is given by the area of the extremal area bulk surface homologous to $A$ plus the bulk entropy in the region between that surface and the boundary \cite{Ryu:2006ef,Ryu:2006bv,Hubeny:2007xt,Faulkner:2013ana,Lewkowycz:2013nqa,Dong:2016hjy}. This formula was shown to hold at $O((1/N)^0)$ in the large-$N$ expansion. In \cite{Engelhardt:2014gca}, Engelhardt and Wall proposed that the all-orders modification of this formula is to replace the extremal area surface with the Quantum Extremal Surface (QES), which is defined as the surface which extremizes the generalized entropy: the surface area plus the entropy in the region between the surface and $A$. Though the Engelhardt-Wall prescription remains unproven, we will assume that it is the correct all-orders prescription for computing the boundary entropy of $A$. We denote the QES of $A$ as $e(A)$.

The entanglement wedge $\mathcal{E}(A)$ is the bulk region spacelike-related to $e(A)$ on the $A$ side of the surface. This is the bulk region believed to be dual to $A$ in subregion duality \cite{Czech:2012bh}. Dong, Harlow and Wall have argued that this is the case using the formalism of quantum error correction \cite{Dong:2016eik,Harlow:2016vwg}.

$\mathcal{C} \subseteq \mathcal{E}$ is the property that \emph{the entanglement wedge $\mathcal{E}(A)$ associated to a boundary region $A$ completely contains the causal wedge \(\mathcal{C}(A)\) associated to $A$}. \(\cse\) can equivalently be formulated as stating that \(e(A)\) is out of causal contact with \(D(A)\), i.e. $e(A) \cap (I^+(D(A) \cup I^-(D(A)) = \emptyset$. In our proofs below we will use this latter characterization.

Subregion duality requires $\mathcal{C} \subseteq \mathcal{E}$ because the bulk region dual to a boundary region $A$ should at least include all of the points that can both send and receive causal signals to and from $D(A)$. Moreover, if $\mathcal{C} \subseteq \mathcal{E}$ were false then it would be possible to use local unitary operators in $D(A)$ to send a bulk signal to $e(A)$ and thus change the entropy associated to the region \cite{Czech:2012bh,Wall:2012uf,Engelhardt:2014gca,Headrick:2014cta}. That is, of course, not acceptable, as the von Neumann entropy is invariant under unitary transformations.

This condition has been discussed at the classical level in \cite{Headrick:2014cta,Wall:2012uf}. In the semiclassical regime, Engelhardt and Wall \cite{Engelhardt:2014gca} have shown that it follows from the generalized second law (GSL) of causal horizons. We will show in Sec. \ref{sec:proofs} that $\mathcal{C} \subseteq \mathcal{E}$ is also implied by Entanglement Wedge Nesting.

\subsubsection*{Entanglement Wedge Nesting (EWN)}
The strongest of the geometrical constraints we consider is EWN. In the framework of subregion duality, EWN is the property that a strictly larger boundary region should be dual to a strictly larger bulk region. More precisely, \emph{for any two boundary regions \(A\) and \(B\) with domain of dependence \(D(A)\) and \(D(B)\) such that \(D(A) \subset D(B)\), we have \(\mathcal{E}(A) \subset \mathcal{E}(B)\)}. 

This property was identified as important for subregion duality and entanglement wedge reconstruction in \cite{Czech:2012bh,Wall:2012uf}, and was proven by Wall at leading order in \(G\hbar\) assuming the null curvature condition \cite{Wall:2012uf}. We we will show in Sec. \ref{sec:proofs} that the Quantum Focussing Conjecture (QFC) \cite{Bousso:2015mna} implies EWN in the semiclassical regime assuming the generalization of HRT advocated in \cite{Engelhardt:2014gca}.

\subsection{Constraints on Semiclassical Quantum Gravity}\label{sec:gloss:inequalities}

Reasonable theories of matter are often assumed to satisfy various energy conditions. The least restrictive of the classical energy conditions is the null energy condition (NEC), which states that
\begin{align}
	T_{kk} \equiv T_{ab} \, k^{a}k^{b} \geq 0~,
\end{align}
for all null vectors \(k^a\).
This condition is sufficient to prove many results in classical gravity. In particular, many proofs hinge on the classical focussing theorem \cite{Wald}, which follows from the NEC and ensures that light-rays are focussed whenever they encounter matter or gravitational radiation:
\begin{align}
	\theta' \equiv \frac{d}{d\lambda} \theta \leq 0~,
\end{align}
where \(\theta\) is the expansion of a null hypersurface and $\lambda$ is an affine parameter.

Quantum fields are known to violate the NEC, and therefore are not guaranteed to focus light-rays. It is desirable to understand what (if any) restrictions on sensible theories exist in quantum gravity, and which of the theorems which rule out pathological phenomenon in the classical regime have quantum generalizations. In the context of AdS/CFT, the NEC guarantees that the bulk dual is consistent with boundary microcausality \cite{Gao:2000ga} and holographic entanglement entropy \cite{Wall:2012uf,Callan:2012ip,Headrick:2013zda,Headrick:2014cta}, among many other things.

In this subsection, we outline three statements in semiclassical quantum gravity which have been used to prove interesting results when the NEC fails. They are presented in order of increasing strength. We will find in sections \ref{sec:energyentropy} and \ref{sec:proofs} that each of them has a unique role to play in the proper functioning of the bulk-boundary duality.

\subsubsection*{Achronal Averaged Null Energy Condition}

The achronal averaged null energy condition (AANEC) \cite{Wald:1991xn} states that
\begin{align}\label{ANECdef}
	\int T_{kk} \, d\lambda \geq 0~,
\end{align}
where the integral is along a complete achronal null curve (often called a ``null line"). Local negative energy density is tolerated as long as it is accompanied by enough positive energy density elsewhere. The \emph{achronal} qualifier is essential for the AANEC to hold in curved spacetimes. For example, the Casimir effect as well as quantum fields on a Schwarzschild background can both violate the ANEC \cite{Klinkhammer:1991ki,Visser:1996iv} for chronal null geodesics. An interesting recent example of violation of the ANEC for chronal geodesics in the context of AdS/CFT was studied in \cite{Gao:2016bin}.

The AANEC is fundamentally a statement about quantum field theory formulated in curved backgrounds containing complete achronal null geodesics. It has been proven for QFTs in flat space from monotonicity of relative entropy \cite{Faulkner:2016mzt}, as well as causality \cite{Hartman:2016lgu}. Roughly speaking, the AANEC ensures that when the backreaction of the quantum fields is included it will focus null geodesics and lead to time delay. This will be made more precise in Sec.~\ref{sec:AANECBCC} when we discuss a proof of the boundary causality condition (BCC) from the AANEC.

\subsubsection*{Generalized Second Law}

The generalized second law (GSL) of horizon thermodynamics states that the generalized entropy (defined below) of a causal horizon cannot decrease in time. 

Let $\Sigma$ denote a Cauchy surface and let $\sigma$ denote some (possibly non-compact) codimension-2 surface dividing $\Sigma$ into two distinct regions. We can compute the von Neumann entropy of the quantum fields on the region outside of $\sigma$, which we will denote $S_{\rm out}$\footnote{The choice of ``outside" is arbitrary. In a globally pure state both sides will have the same entropy, so it will not matter which is the ``outside." In a mixed state the entropies on the two sides will not be the same, and thus there will be two generalized entropies associated to the same surface. The GSL, and all other properties of generalized entropy, should apply equally well to both.}. The generalized entropy of this region is defined to be
\begin{equation}\label{SgenDef}
S_{\rm gen} = S_{\rm grav} + S_{\rm out}
\end{equation}
where \(S_{\rm grav}\) is the geometrical/gravitational entropy which depends on the theory of gravity. For Einstein gravity, it is the familiar Bekenstein-Hawking entropy. There will also be Planck-scale suppressed corrections\footnote{There will also be stringy corrections suppressed by $\alpha'$. As long as we are away from the stringy regime, these corrections will be suppressed in a way that is similar to the Planck-suppressed ones, and so we will not separately track them.}, denoted \(Q\), such that it has the general form
\begin{align}
	S_{\rm grav} = \frac{A}{4G\hbar} + Q
\end{align}
There is mounting evidence that the generalized entropy is finite and well-defined in perturbative quantum gravity, even though the split between matter and gravitational entropy depends on renormalization scale. See the appendix of \cite{Bousso:2015mna} for details and references.

The \emph{quantum expansion} \(\Theta\) can be defined (as a generalization of the classical expansion \(\theta\)) as the functional derivative per unit area of the generalized entropy along a null congruence \cite{Bousso:2015mna}:
\begin{align}\label{ThetaDef}
	\Theta[\sigma(y); y] &\equiv \frac{4G\hbar}{\sqrt{h}} \frac{\delta S_{\rm gen}}{\delta \sigma(y)} \\ 
    &= \theta + \frac{4G\hbar}{\sqrt{h}} \frac{\delta Q}{\delta \sigma(y)}+ \frac{4G\hbar}{\sqrt{h}} \frac{\delta S_{\rm out}}{\delta \sigma(y)}
\end{align}
where $\sqrt{h}$ denotes the determinant of the induced metric on $\sigma$, which is parametrized by \(y\). These functional derivatives denote the infinitesimal change in a quantity under deformations of the surface at coordinate location $y$ along the chosen null congruence. To lighten the notation, we will often omit the argument of \(\Theta\).

A future (past) causal horizon is the boundary of the past (future) of any future-infinite (past-infinite) causal curve \cite{Jacobson:2003wv}. For example, in an asymptotically AdS spacetime any collection of points on the conformal boundary defines a future and past causal horizon in the bulk. The generalized second law (GSL) is the statement that the quantum expansion is always nonnegative towards the future on any future causal horizon
\begin{align}\label{GSL}
	\Theta \geq 0~,
\end{align}
with an analogous statement for a past causal horizon.

In the semiclassical \(G\hbar \to 0\) limit, Eq.~\eqref{ThetaDef} reduces to the classical expansion \(\theta\) if it is nonzero, and the GSL becomes the Hawking area theorem \cite{Hawking:1971tu}. The area theorem follows from the NEC. 

Assuming the validity of the GSL allows one to prove a number of important results in semiclassical quantum gravity \cite{C:2013uza,Engelhardt:2014gca}. In particular, Wall has shown that it implies the AANEC \cite{Wall:2009wi}, as we will review in Section \ref{sec:energyentropy}, and $\mathcal{C} \subseteq \mathcal{E}$ \cite{Engelhardt:2014gca}, reviewed in Section \ref{sec:proofs} (see Fig.~\ref{fig:CHART}).

\subsubsection*{Quantum Focussing Conjecture}

The Quantum Focussing Conjecture (QFC) was conjectured in \cite{Bousso:2015mna} as a quantum generalization of the classical focussing theorem, which unifies the Bousso Bound and the GSL. The QFC states that the functional derivative of the quantum expansion along a null congruence is nowhere increasing:
\begin{equation}\label{QFC}
\frac{\delta \Theta[\sigma(y_1);y_1]}{\delta \sigma(y_2)}\leq 0~.
\end{equation}
In this equation, \(y_1\) and \(y_2\) are arbitrary. When \(y_1 \neq y_2\), only the \(S_{\rm out}\) part contributes, and the QFC follows from strong subadditivity of entropy \cite{Bousso:2015mna}. For notational convenience, we will often denote the ``local" part of the QFC, where \(y_1 = y_2\), as\footnote{Strictly speaking, we should factor out a delta function $\delta(y_1-y_2)$ when discussing the local part of the QFC \cite{Bousso:2015wca,Koeller:2015qmn}. Since the details of this definition are not important for us, we will omit this in our notation.}
\begin{align}
	\Theta'[\sigma(y);y] \leq 0.
\end{align}
Note that while the GSL is a statement only about causal horizons, the QFC is conjectured to hold on any cut of any null hypersurface.

If true, the QFC has several non-trivial consequences which can be teased apart by applying it to different null surfaces \cite{Bousso:2015mna,Bousso:2015eda,Engelhardt:2014gca}. In Sec. \ref{sec:proofs} we will see that EWN can be added to this list.

\subsubsection*{Quantum Null Energy Condition}
When applied to a locally stationary null congruence, the QFC leads to the Quantum Null Energy Condition (QNEC) \cite{Bousso:2015mna,Koeller:2015qmn}. Applying the Raychaudhuri equation and Eqs.~\eqref{SgenDef}, \eqref{ThetaDef} to the statement of the QFC \eqref{QFC}, we find
\begin{align}
	0 \geq \Theta' = -\,\frac{\theta^2}{D-2} - \sigma^2 - 8 \pi G\, T_{kk} + \frac{4 G \hbar}{\sqrt{h}} \left( S''_{\rm out} - S'_{\rm out} \theta \right)
\end{align}
where \(S''_{\rm out}\) is the local functional derivative of the matter entropy to one side of the cut. If we consider a locally stationary null hypersurface satisfying \(\theta^2 = \sigma^2 = 0\) in a small neighborhood, this inequality reduces to the statement of the \emph{Quantum Null Energy Condition} (QNEC) \cite{Bousso:2015mna}:
\begin{align}\label{QNECdef}
	T_{kk} \geq \frac{\hbar}{2\pi\sqrt{h}} \,S''_{\rm out}
\end{align}
It is important to notice that the gravitational coupling \(G\) has dropped out of this equation. The QNEC is a statement purely in quantum field theory which can be proven or disproven using QFT techniques. It has been proven for both free fields \cite{Bousso:2015wca} and holographic field theories at leading order in \(G\hbar\) \cite{Koeller:2015qmn}.\footnote{There is also evidence \cite{Fu:2016avb} that the QNEC holds in holographic theories where the entropy is taken to be the casual holographic information \cite{Hubeny:2012wa}, instead of the von Neumann entropy.} In Section \ref{sec:proofs} of this paper, we generalize the holographic proof to all orders in \(G\hbar\). These proofs strongly suggest that the QNEC is a true property of general quantum field theories.\footnote{The free-field proof of \cite{Bousso:2015wca} was for arbitrary cuts of Killing horizons. The holographic proof of \cite{Koeller:2015qmn} (generalized in this paper) showed the QNEC for a locally stationary (\(\theta = \sigma = 0\)) portion of \emph{any} Cauchy-splitting null hypersurface in flat space.} In the classical \(\hbar \to 0\) limit, the QNEC becomes the NEC.

\subsubsection*{Quantum Half-Averaged Null Energy Condition}

The quantum half-averaged energy condition (QHANEC) is an inequality involving the integrated stress tensor and the first null derivative of the entropy on one side of any locally-stationary Cauchy-splitting surface subject to a causality condition (described below):
\begin{align}\label{eqn:QHANEC}
	\int_{\lambda}^{\infty} T_{kk} \,d\tilde\lambda \geq -\frac{\hbar}{2\pi\sqrt{h}} S'(\lambda),
\end{align}
Here \(k^a\) generates a null congruence with vanishing expansion and shear in a neighborhood of the geodesic and $\lambda$ is the affine paramter along the geodesic. The geodesic thus must be of infinite extent and have \(R_{ab}k^a k^b = C_{abcd}k^a k^c = 0\) everywhere along it. The aforementioned causality condition is that the Cauchy-splitting surfaces used to define $S(\lambda)$ should not be timelike-related to the half of the null geodesic $T_{kk}$ is integrated over. Equivalently, $S(\lambda)$ should be well-defined for all $\lambda$ from the starting point of integration all the way to $\lambda=\infty$.

The causality condition and the stipulation that the null geodesic in (\ref{eqn:QHANEC}) be contained in a locally stationary congruence ensures that the QHANEC follows immediately from integrating the QNEC (Eq.~\eqref{QNECdef}) from infinity (as long as the entropy isn't evolving at infinite affine parameter, i.e., $S'(\infty)=0$). Because the causality condition is a restriction on the global shape of the surface, there will be situations where the QNEC holds locally but we cannot integrate to arrive at a QHANEC.

The QHANEC appears to have a very close relationship to monotonicity of relative entropy. Suppose that the modular Hamiltonian of the portion of a null plane above an arbitrary cut \(u= \sigma(y)\) (where $u$ is a null coordinate) is given by
\begin{equation}\label{eqn:modham}
K[\sigma(y)] = \int d^{d-2}y \int_{\sigma(y)}^\infty d\lambda~(\lambda - \sigma(y))\, T_{kk}
\end{equation}
Then (\ref{eqn:MRE}) becomes monotonicity of relative entropy. As of yet, there is no known general proof in the literature of \eqref{eqn:modham}, though for free theories it follows from the enhanced symmetries of null surface quantization \cite{Wall:2011hj}. Eq.~\eqref{eqn:modham} can be also be derived for holographic field theories \cite{Arvin:FutureWork}. It has also been shown that linearized backreaction from quantum fields obeying the QHANEC will lead to a spacetime satisfying the GSL \cite{Wall:2011hj}.\footnote{It has been shown \cite{Bunting:2015sfa} that holographic theories also obey the QHANEC when the causal holographic information \cite{Hubeny:2012wa} is used, instead of the von Neumann entropy. This implies a second law for the causal holographic information in holographic theories.}

In Sec.~\ref{sec:proofs}, we will find that $C \subseteq E$ implies the QHANEC on the boundary.

\section{Relationships Between Entropy and Energy Inequalities}\label{sec:energyentropy}
The inequalities discussed in the previous section are not all independent. In this section we discuss the logical relationships between them.

\subsubsection*{GSL implies AANEC}\label{gslAANEC}

Wall has shown \cite{Wall:2009wi} that the GSL implies the AANEC in spacetimes which are linearized perturbations of classical backgrounds, where the classical background  obeys the null energy condition (NEC). Here, we point out that this proof is sufficient to prove the AANEC from the GSL in the semi-classical regime, to all orders in \(G\hbar\) (see Sec.~\ref{sec:glossary}).

Because the AANEC is an inequality, in the semi-classical \(G\hbar \to 0\) limit its validity is determined by the leading non-zero term in the \(G\hbar\) expansion. Suppose that this term is order \((G\hbar)^m\). Suppose also that at order \((G\hbar)^{m-1}\) the metric contains a complete achronal null geodesic \(\gamma\), i.e. a null geodesic without a pair of conjugate points. (If at this order no such geodesics exist, the AANEC holds trivially at this order as well as all higher orders, as higher-order contributions to the metric cannot make a chronal geodesic achronal.) Achronality guarantees that \(\gamma\) lies in both a future and past causal horizon, \(\mathcal{H}^\pm\).

Wall's proof required that, in the background spacetime, the expansion and shear vanish along \(\gamma\) in both \(\mathcal{H}^+\) and \(\mathcal{H}^-\). Wall used the NEC in the background spacetime to derive this, but here we note that the NEC is not necessary given our other assumptions. The ``background" for us is the \(O((G\hbar)^{m-1})\) part of the metric. Consider first the past causal horizon, \(\mathcal{H}^-\), which must satisfy the boundary condition \(\theta(-\infty) \to 0\). Since \(\gamma\) is achronal, the expansion \(\theta\) of \(\mathcal{H}^-\) cannot blow up to \(-\infty\) anywhere along $\gamma$. As $\lambda \to \infty$, $\theta$ can either remain finite or blow up in the limit. Suppose first that \(\theta\) asymptotes to a finite constant as \(\lambda \to \infty\). Then \(\lim_{\lambda \to \infty} \theta' = 0\). Assuming the matter stress tensor dies off at infinity (as it must for the AANEC to be well-defined), Raychaudhuri's equation gives \(\lim_{\lambda \to \infty} \theta' = -\theta^2/(D-2) - \sigma^2\), the only solution to which is\footnote{We absorb the graviton contribution to the shear into the stress tensor.}
\begin{align}
	\lim_{\lambda \to \infty} \theta = \lim_{\lambda \to \infty}\sigma = 0~.
\end{align}
Similar arguments apply to \(\mathcal{H}^+\). This also implies that \(\mathcal{H}^+ = \mathcal{H}^-\). The rest of the proof follows \cite{Wall:2009wi}. This proves the AANEC at order \((G\hbar)^m\).

The alternative case is that $|\theta| \to \infty$ as $\lambda \to \infty$. But if $T_{kk}$ dies off at infinity, then for large enough $\lambda$ we have $\theta' < -\theta^2/(D-2) + \epsilon$ for some $\epsilon$. Then a simple modification of the standard focussing argument shows that $\theta$ goes to $-\infty$ at finite affine parameter, which is a contradiction.

\subsubsection*{QFC implies GSL}
In a manner exactly analogous to the proof of the area theorem from classical focusing, the QFC can be applied to a causal horizon to derive the GSL. Consider integrating  Eq.~\ref{QFC} from future infinity along a generator of a past causal horizon:
\begin{equation}\label{eqn:IntQFC}
\int d^{d-2}y \, \sqrt{h}\, \int_{\lambda}^{\infty}  d\tilde\lambda \,\Theta^{\prime}[\sigma(y,\tilde\lambda);y] \leq 0
\end{equation}
Along a future causal horizon, \(\theta \to 0\) as \(\lambda \to \infty\), and it is reasonable to expect the matter entropy \(S_{\rm out}\) to stop evolving as well. Thus \(\Theta \to 0\) as \(\lambda \to \infty\), and the integrated QFC then trivially becomes 
\begin{equation}\label{eqn:GSL}
\Theta[\sigma(y);y] \geq 0
\end{equation}
which is the GSL.

\subsubsection*{QHANEC implies AANEC}
In flat space, all achronal null geodesics lie on a null plane. Applying the QHANEC to cuts of this null plane taking \(\lambda \to -\infty\) produces the AANEC, Eq.~\eqref{ANECdef}.

\section{Relationships Between Entropy and Energy Inequalities and Geometric Constraints}\label{sec:proofs}
In this section, we discuss how the bulk generalized entropy conditions reviewed in Sec.~\ref{sec:gloss:inequalities} imply the geometric conditions EWN, $\mathcal{C} \subseteq \mathcal{E}$ and BCC (described in Sec.~\ref{sec:gloss:geometrical}). We also explain how these geometric conditions imply the boundary QNEC, QHANEC and AANEC.

\subsection{EWN implies $\mathcal{C} \subseteq \mathcal{E}$ implies the BCC}
\subsubsection*{EWN implies $\mathcal{C} \subseteq \mathcal{E}$}

\begin{figure}\label{fig:EWN_implies_EWCC_fig}
	\centering
	\includegraphics[width= 5cm]{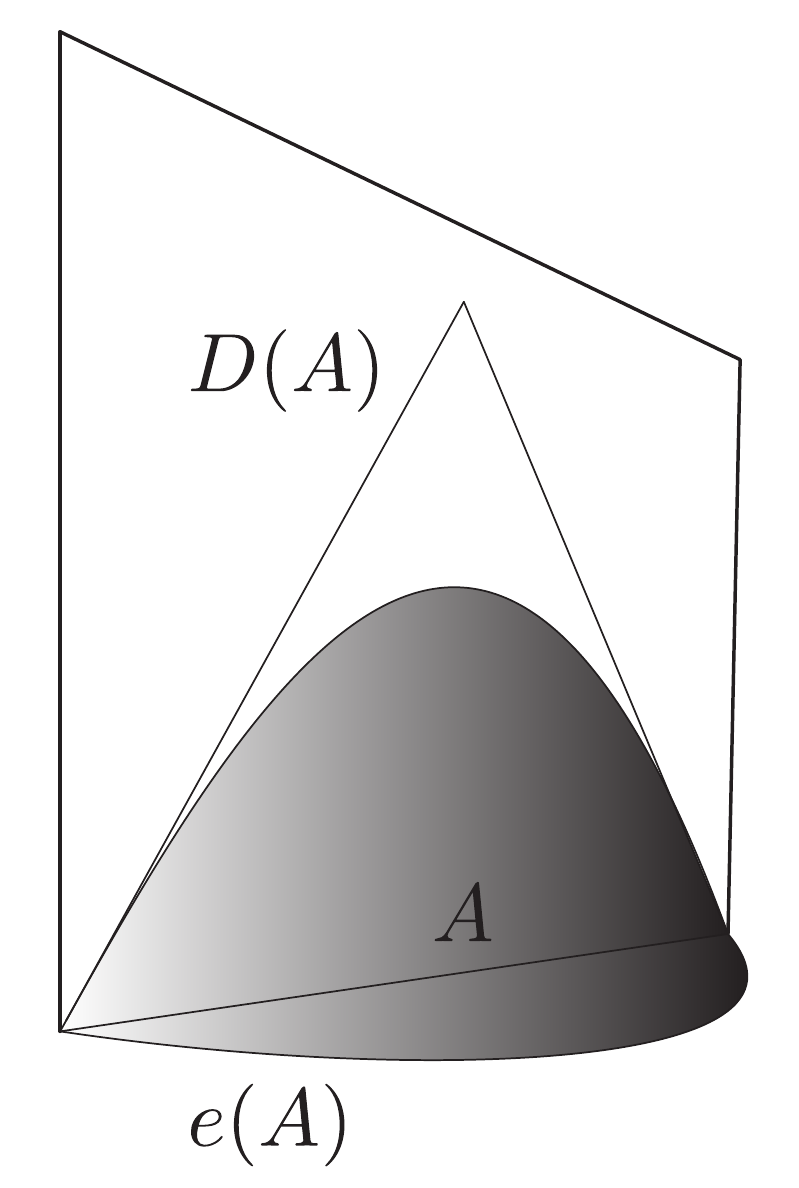}
	\caption{The causal relationship between $e(A)$ and $D(A)$ is pictured in an example spacetime that violates $\mathcal{C} \subseteq \mathcal{E}$. The boundary of $A$'s entanglement wedge is shaded. Notably, in $\mathcal{C}\subseteq \mathcal{E}$ violating spacetimes, there is necessarily a portion of $D(A)$ that is timelike related to $e(A)$. Extremal surfaces of boundary regions from this portion of $D(A)$ are necessarily timelike related to $e(A)$, which violates EWN.}
\label{fig:EWN_implies_EWCC_fig}
\end{figure}

The \(\cse\) and EWN conditions were defined in Sec.~\ref{sec:gloss:geometrical}. There, we noted that \(\cse\) can be phrased as the condition that the extremal surface \(e(A)\) for some boundary region \(A\) lies outside of timelike contact with \(D(A)\). We will now prove that EWN implies \(\cse\) by proving the contrapositive: we will show that if \(\cse\) is violated, there exist two boundary regions \(A,B\) with nested domains of dependence, but whose entanglement wedges are not nested.

Consider an arbitrary region $A$ on the boundary. $\mathcal{C} \subseteq \mathcal{E}$ is violated if and only if there exists at least one point $p \in e(A)$ such that $p \in I^+(D(A)) \cup I^-(D(A))$, where \(I^+\) (\(I^-\)) denotes the chronological future (past). Then violation of $\cse$ is equivalent to the exisence of a timelike curve connecting $e(A)$ to $D(A)$. Because \(I^{+}\) and \(I^{-}\) are open sets, there exists an open neighborhood $\mathcal{O} \subset D(A)$ such that every point of $\mathcal{O}$ is timelike related to $e(A)$ (see Figure \ref{fig:EWN_implies_EWCC_fig}). Consider a new boundary region $B \subset \mathcal{O}$. Again by the openness of $I^{+}$ and $I^{-}$, the corresponding entanglement wedge \(\mathcal{E}(B)\) also necessarily contains points that are timelike related to $e(A)$. Since \(\mathcal{E}(A)\) is defined to be all points \emph{spacelike-related} to \(e(A)\) on the side towards \(A\), \(\mathcal{E}(B) \nsubseteq \mathcal{E}(A)\). But by construction \(D(B) \subseteq D(A)\), and thus EWN is violated.

In light of this argument, we have an additional characterization of the condition $\cse$: $\cse$ is what guarantees that $\mathcal{E}(A)$ contains $D(A)$, which is certainly required for consistency of bulk reconstruction.

\subsubsection*{$\mathcal{C} \subseteq \mathcal{E}$ implies the BCC}

\begin{figure}
	\centering
	\includegraphics[width= 8 cm]{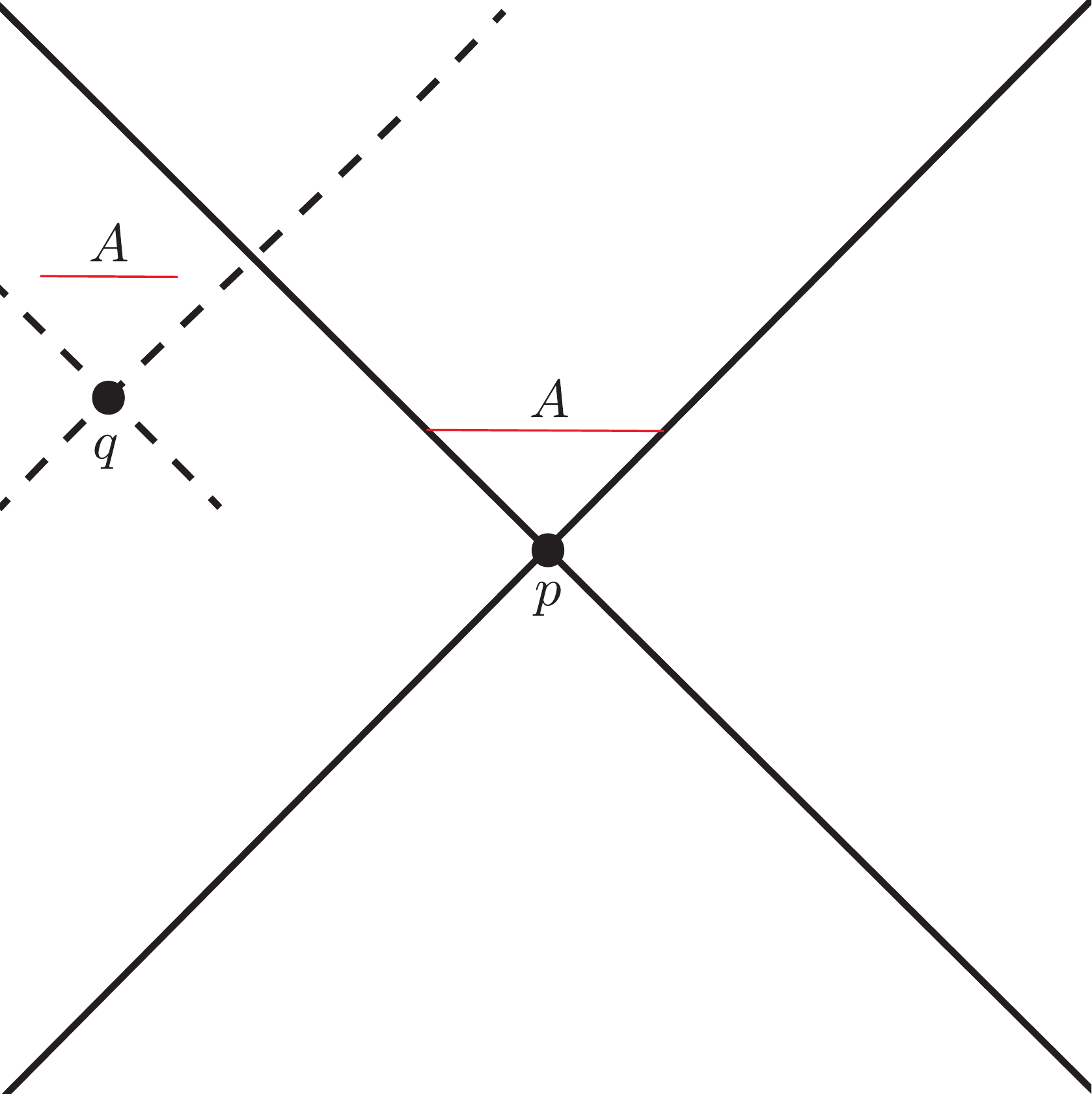}
	\caption{The boundary of a BCC-violating spacetime is depicted, which gives rise to a violation of $\mathcal{C} \subseteq \mathcal{E}$. The points $p$ and $q$ are connected by a null geodesic through the bulk. The boundary of $p$'s lightcone with respect to the AdS boundary causal structure is depicted with solid black lines. Part of the boundary of $q$'s lightcone is shown with dashed lines. The disconnected region $A$ is defined to have part of its boundary in the timelike future of $q$ while also satisfying $p \in D(A)$. It follows that $e(A)$ will be timelike related to $D(A)$ through the bulk, violating $\mathcal{C} \subseteq \mathcal{E}$.}
	\label{fig:C_EimpliesBCC}
\end{figure}

We prove the contrapositive. If the BCC is violated, then there exists a bulk null geodesic from some boundary point $p$ that returns to the boundary at a point $q$ not to the future of $p$ with respect to the boundary causal structure. Therefore there exist points in the timelike future of $q$ that are also not to the future of $p$. 

If $q$ is not causally related to $p$ with respect to the boundary causal structure, we derive a contradiction as follows. Define a boundary subregion $A$ with two disconnected parts: one that lies entirely within the timelike future of $q$ but outside the future of $p$, and one composed of all the points in the future lightcone of $p$ on a boundary timeslice sufficiently close to $p$ such that $A$ is completely achronal. By construction, $p \in D(A)$. Moreover, because $\partial A$ includes points timelike related to $q$, $e(A)$ includes points timelike related to $q$ and by extension $p$. Therefore $A$ is an achronal boundary subregion whose extremal surface contains points that are timelike related to $D(A)$. See Figure \ref{fig:C_EimpliesBCC}.

If $q$ is in the past of $p$, then a contradiction is reached more easily. Define a boundary subregion $A$ as the intersection of $p$'s future lightcone with any constant time slice sufficiently close to $p$, chosen so that $e(A)$ is not empty. Then $p$ is in $D(A)$ and $q$ is in $I^-(A)$ according to the boundary causal structure (though according to the bulk causal structure it is in $J^+(p)$). Hence $e(A)$ is timelike related to $D(A)$ in the bulk causal structure, which is the sought-after contradiction.


\subsection{Semiclassical Quantum Gravity Constraints Imply Geometric Constraints}
\subsubsection*{Quantum Focussing implies Entanglement Wedge Nesting}

Consider a boundary region $A$ associated with boundary domain of dependence $D(A)$. As above, we denote the quantum extremal surface anchored to $\partial A$ as $e(A)$. For any other boundary region, $B$, such that $D(B) \subset D(A)$, we will show that $\mathcal{E}(B) \subset \mathcal{E}(A)$, assuming the QFC.

Since we are treating quantum corrections perturbatively, every quantum extremal surface is located near a classical extremal area surface.\footnote{Another possibility is that quantum extremal surfaces which exist at finite $G\hbar$ move off to infinity as $G\hbar\to 0$. In that case there would be no associated classical extremal surface. If we believe that the classical limit is well-behaved, then these surfaces must always be subdominant in the small $G\hbar$ limit, and so we can safely ignore them.} Wall proved in \cite{Wall:2012uf} that $\mathcal{E}(B) \subset \mathcal{E}(A)$ is true at the classical level if we assume classical focussing. Thus to prove the quantum statement within perturbation theory we only need to consider those (nongeneric) cases where $e(B)$ happens to intersect the boundary of $\mathcal{E}(A)$ classically.\footnote{The only example of this that we are aware of is in vacuum AdS where $A$ is the interior of a sphere on the boundary and $B$ is obtained by deforming a portion of the sphere in an orthogonal null direction.} In such a case, one might worry that a perturbative quantum correction could cause $e(B)$ to exit $\mathcal{E}(A)$. We will now argue that this does not happen.\footnote{For now we ignore the possibility of phase transitions. They will be treated separately below.} 

First, deform the region $B$ slightly to a new region $B' \subset A$ such that $e(B')$ lies within $\mathcal{E}(A)$ classically. Then, since perturbative corrections cannot change this fact, we will have $\mathcal{E}(B') \subset \mathcal{E}(A)$ even at the quantum level. Now, following \cite{Engelhardt:2014gca}, we show that in deforming $B'$ back to $B$ we maintain EWN.

The QFC implies that the null congruence generating the boundary of $I^\pm(e(A))$ satisfies $\dot{\Theta} \leq 0$. Combined with $\Theta = 0$ at $e(A)$ (from the definition of quantum extremal surface), this implies that every point on the boundary of $\mathcal{E}(A)$ satisfies $\Theta \le 0$. Therefore the boundary of $\mathcal{E}(A)$ is a quantum extremal barrier as defined in \cite{Engelhardt:2014gca}, and so no continuous family of quantum extremal surfaces can cross the boundary of $\mathcal{E}(A)$. Thus, as we deform $B'$ back into $B$, the quantum extremal surface is forbidden from exiting $\mathcal{E}(A)$. Therefore $e(B) \subset \mathcal{E}(A)$, and by extension $\mathcal{E}(B) \subset \mathcal{E}(A)$.

Finally we will take care of the possibility of a phase transition. A phase transition occurs when there are multiple quantum extremal surfaces for each region, and the identity of the one with minimal generalized entropy switches as we move within the space of regions. This causes the entanglement wedge to jump discontinuously, and if it jumps the ``wrong way" then EWN could be violated. Already this would be a concern at the classical level, but it was shown in \cite{Wall:2012uf} that classically EWN is always obeyed even accounting for the possibility of phase transitions. So we only need to convince ourselves that perturbative quantum corrections cannot change this fact.

Consider the infinite-dimensional parameter space of boundary regions. A family of quantum (classical) extremal surfaces determines a function on this parameter space given by the generalized entropy (area) of the extremal surfaces. A phase transition occurs when two families of extremal surfaces have equal generalized entropy (or area), and is associated with a codimension-one manifold in parameter space. In going from the purely classical situation to the perturbative quantum situation, two things will happen. First, the location of the codimension-one phase transition manifold in parameter space will be shifted. Second, within each family of extremal surfaces, the bulk locations of the surfaces will be perturbatively shifted. We can treat these two effects separately. 

In the vicinity of the phase transition (in parameter space), the two families of surfaces will be classically separated in the bulk and classically obey EWN, as proved in \cite{Wall:2012uf}. A perturbative shift in the parameter space location of the phase transition will not change whether EWN is satisfied classically. That is, the {\em classical} surfaces in each extremal family associated with the neighborhood of {\em quantum} phase transition will still be separated classically in the bulk. Then we can shift the bulk locations of the classical extremal surfaces to the quantum extremal surfaces, and since the shift is only perturbative there is no danger of introducing a violation of EWN.

It would be desirable to have a more unified approach to this proof in the quantum case that does not rely so heavily on perturbative arguments. We believe that such an approach is possible, and in future work we hope to lift all of the results of  \cite{Wall:2012uf} to the quantum case by the replacement of ``area" with ``generalized entropy" without having to rely on a perturbative treatment. 

\subsubsection*{Generalized Second Law implies $\mathcal{C} \subseteq \mathcal{E}$}
This proof can be found in \cite{Engelhardt:2014gca}, but we elaborate on it here to illustrate similarities between this proof and the proof that QFC implies EWN. 

\paragraph{Wall's Lemma}\label{lemma:Walls}
We remind the reader of a fact proved as Theorem 4 in \cite{Wall:2012uf}.\footnote{Wall's Lemma is a significant part of the extremal surface barriers argument in \cite{Engelhardt:2014gca}.} Let two boundary anchored co-dimension two, spacelike surfaces $M$ and $N$, which contain the point $ p \in M \cap N$ such that they are also tangent at $p$. Both surfaces are Cauchy-splitting in the bulk. Suppose that $M$ lies completely to one side of $N$. In the classical regime, Wall shows that either there exists some point $x$ in a neighborhood of $p$ where 
\begin{equation}\label{eqn:thetabound}
\theta_N(x) > \theta_M(x)
\end{equation}
or the two surfaces agree everywhere in the neighborhood. These expansions are associated to the exterior facing, future null normal direction. 

In the semi-classical regime, this result can be improved to bound the quantum expansions
\begin{equation}\label{eqn:genbound}
\Theta_1(x) > \Theta_2(x)
\end{equation}
where $x$ is some point in a neighborhood of $p$. The proof of this quantum result requires the use of strong sub-additivity, and works even when bulk loops generate higher derivative corrections to the generalized entropy \cite{C:2013uza}.

\begin{figure}
	\centering
	\includegraphics[width= 9cm]{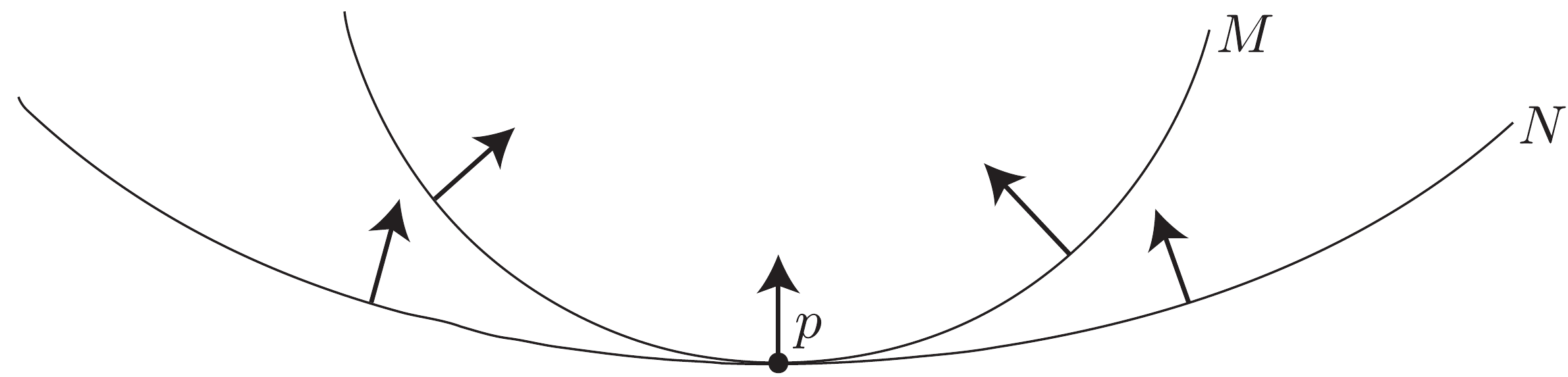}
	\caption{The surface $M$ and $N$ are shown touching at a point $p$. In this case, $\theta_M < \theta_N$. The arrows illustrate the projection of the null orthogonal vectors onto the Cauchy surface.}
	\label{fig:touching}
\end{figure}

We now proceed to prove $\cse$ from the GSL by contradiction. Suppose that the causal wedge lies at least partly outside the entanglement wedge. In this discussion, by the ``boundary of the causal wedge," we mean the intersection of the past of $I^-(\partial D(A))$ with the Cauchy surface on which $e(A)$ lies. Consider continuously shrinking the boundary region associated to the causal wedge. The causal wedge will shrink continuously under this deformation. At some point, $\mathcal{C}(A)$ must shrink inside $\mathcal{E}(A)$. There exists some Cauchy surface such that its intersection with the boundary of the causal wedge touches the original extremal surface as depicted in Figure \ref{fig:touching}. There, $M$ is the intersection of the boundary of the causal wedge of the shrunken region with the Cauchy surface and $N$ is $e(A)$.

Assuming genericity of the state, the two surfaces cannot agree in this neighborhood. At this point, by the above lemma, the quantum expansions should obey 
\begin{equation}
\Theta_{e}(x)>\Theta_{c}(x)
\end{equation}
for $x$ in some neighborhood of $p$.  The Wall-Engelhardt prescription tells us that the entanglement wedge boundary should be given by the quantum extremal surface \cite{Engelhardt:2014gca} and so
\begin{equation}
\Theta_e(x) = 0 > \Theta_c(x)
\end{equation}
Thus, the GSL is violated at some point along this causal surface, which draws the contradiction.

\subsubsection*{AANEC implies Boundary Causality Condition}\label{sec:AANECBCC}

The Gao-Wald proof of the BCC \cite{Gao:2000ga} uses the fact --- which follows from their assumptions of the NEC and null generic condition (discussed below) --- that all complete null geodesics through the bulk contain a pair of conjugate points.\footnote{Intuitively speaking, points \(p\) and \(q\) along a geodesic \(\gamma\) are conjugate if an infinitesimally nearby geodesic intersects \(\gamma\) at both \(p\) and \(q\). This can be shown to be equivalent to the statement that the expansion of a congruence through \(p\) approaches \(-\infty\) at \(q\). See e.g. \cite{Wald} for details.} Here, we sketch a slight modification of the proof which instead assumes the achronal averaged null energy condition (AANEC).

We prove that the AANEC implies BCC by contradiction. Let the spacetime satisfy the null generic condition \cite{Wald}, so that each null geodesic encounters at least some matter or gravitational radiation.\footnote{Mathematically, each complete null geodesic should contain a point where \(k^a k^b k_{[c} R_{d]ab[e} k_{f]} \neq 0\).} Violation of the BCC implies that the "fastest" null geodesic \(\gamma\) between two boundary points \(p\) and \(q\)  lies in the bulk. Such a null geodesic is necessarily complete and achronal, as \(p\) and \(q\) are not timelike related. As explained in Sec.~\ref{gslAANEC}, an achronal null geodesic in an AANEC satisfying spacetime is contained in a congruence that is both a past and future causal horizon. Integrating Raychaudhuri's equation along the entire geodesic then gives $0 \leq - \int (\theta^2 + \sigma^2) $, which implies $\theta = \sigma = 0$ everywhere along the geodesic, and hence $\theta' = 0$. Raychaudhuri's equation then says $\theta' = -T_{kk} = 0$ everywhere, which contradicts the generic condition.

\subsection{Geometric Constraints Imply Field Theory Constraints}\label{sec:geoimpFT}
The geometrical constraints EWN, \(\cse\), and BCC have non-trivial implications for the boundary theory. We derive them in this section, which proves the three implications connecting columns two and three of Fig.~\ref{fig:CHART}. The key idea behind all three proofs is the same: express the geometrical constraints in terms of bulk quantities near the asymptotic boundary, and then use near-boundary expansions of the metric and extremal surfaces to convert them into field theory statements.

\subsubsection*{Entanglement Wedge Nesting implies the Boundary QNEC}
At leading order in $G\hbar \sim 1/N$, this proof is the central result of \cite{Koeller:2015qmn}. There the boundary entropy was assumed to be given by the RT formula without the bulk entropy corrections. We give a proof here of how the $1/N$ corrections can be incorporated naturally. We will now show, in a manner exactly analogous to that laid out in \cite{Koeller:2015qmn}, that EWN implies the boundary QNEC. In what follows, we will notice that in order to recover the boundary QNEC, we must use the \textit{quantum extremal surface}, not just the RT surface with FLM corrections \cite{Engelhardt:2014gca}. 

The quantum extremal surface (QES) prescription, as first introduced in \cite{Engelhardt:2014gca}, is the following. To find the entropy of a region $A$ in the boundary theory, first find the minimal codimension-2 bulk surface homologous to \(A\), \(e(A)\), that extremizes the bulk generalized entropy on the side of \(A\). The entropy of $A$ is then given by
\begin{equation}\label{eqn:QES}
S_{\rm A}= S_{\rm gen}(e(A)) = \frac{A_{\rm QES}}{4G\hbar} + S_{\rm bulk}
\end{equation}
Entanglement Wedge Nesting then becomes a statement about how the quantum extremal surface moves under deformations of the boundary region. In particular, for null deformations of the boundary region, EWN states that $e(A)$ moves in a spacelike (or null) fashion.

To state this more precisely, we can set up a null orthogonal basis about $e(A)$. Let $k^{\mu}$ be the inward-facing, future null orthogonal vector along the quantum extremal surface. Let $\ell ^{\mu}$ be its past facing partner with $\ell \cdot k = 1$. Following the prescrition in \cite{Koeller:2015qmn}, we denote the locally orthogonal deviation vector of the quantum extremal surface by $s^{\mu}$. This vector can be expanded in the local null basis as 
\begin{equation}\label{eqn:alphabeta}
s = \alpha k + \beta \ell
\end{equation}
The statement of entanglement wedge nesting then just becomes the statement that $\beta \geq 0$.

In order to find how $\beta$ relates to the boundary QNEC, we would like to find its relation to the entropy. We start by examining the expansion of the extremal surface solution in Fefferman-Graham coordinates. Note that the quantum extremal surface obeys an equation of motion including the bulk entropy term as a source
\begin{equation}\label{eqn:modifiedHRT}
K_{\mu} = -\frac{4G\hbar}{\sqrt{H}}\frac{\delta S_{\rm bulk}}{\delta X^{\mu}}
\end{equation}
Here, $K^{\mu} = \theta_k \ell^{\mu}+ \theta_{\ell} k^{\mu}$ is the extrinsic curvature of the QES. As discussed in \cite{Koeller:2015qmn}, solutions to (\ref{eqn:modifiedHRT}) without the bulk source take the form
\begin{equation}\label{eqn:sourcelesssoln}
\bar{X}^i_{\rm HRT}(y^a,z) = X^i(y^a)+\frac{1}{2(d-2)}z^2K^i(y^a)+...+\frac{z^d}{d}(V^i(y^a)+W^i(y^a)\log z) + o(z^d)
\end{equation}
We now claim that the terms lower order than $z^d$ are unaffected by the presence of the source.  More precisely
\begin{equation}\label{eqn:XQES}
\bar{X}_{\rm QES}^i(y^a,z) = X^i(y^a) + \frac{1}{2(d-2)}z^2 K^i(y^a) +... + \frac{z^d}{d}(V^i_{\rm QES} + W^i(y^a)\log z) + o(z^d)
\end{equation}

This expansion is found by examining the leading order pieces of the extremal surface equation. First, expand (\ref{eqn:modifiedHRT}) to derive
\begin{equation}\label{eqn:expandedHRT}
z^{d-1}\partial_z \left(z^{1-d} f\sqrt{\bar{h}} \bar{h}^{zz}\partial_z \bar{X}^i\right) + \partial_a \left(\sqrt{\bar{h}_{ab}}\bar{h}^{ab} f \partial_b \bar{X}^i \right) = -z^{d-1}4G\hbar f \frac{\delta S_{bulk}}{\delta \bar{X}^j}g^{ji}
\end{equation}
Here we are parameterizing the near-boundary AdS metric in Fefferman-Graham coordinates by 
\begin{align}\label{eqn:AdSmetric}
	ds^2 &= \frac{1}{z^2}\left(dz^2 + g_{ij} dx^i dx^j \right) \\
	&= \frac{1}{z^2}\left(dz^2 + \left[ f(z) \eta_{ij} + \frac{16\pi G_N}{d}z^d t_{ij}\right]dx^idx^j + o(z^d)\right).
\end{align}
The function $f(z)$ encodes the possibility of relevant deformations in the field theory which would cause the vacuum state to differ from pure AdS. Here we have set $L_{AdS} = 1$.

One then plugs in (\ref{eqn:XQES}) to (\ref{eqn:expandedHRT}) to see that the terms lower order than $z^d$ remain unaffected by the presence of the bulk entropy source as long as $\delta S_{\rm bulk}/\delta X^i$ remains finite at $z=0$. We discuss the plausibility of this boundary condition at the end of this section.
 
For null deformations to locally stationary surfaces on the boundary, one can show using \eqref{eqn:XQES} that the leading order piece of $\beta$ in the Fefferman-Graham expansion is order $z^{d-2}$. Writing the coordinates of the boundary entangling surface as a function of some deformation parameter - $X^i(\lambda)$ - we find that \cite{Koeller:2015qmn},
\begin{equation}\label{eqn:preQNEC}
\beta \propto z^{d-2}\left(T_{kk} + \frac{1}{8\pi G_N}k_i \partial_{\lambda}V_{\rm QES}^i\right).
\end{equation}
 
We will now show that $V_{\rm QES}^i$ is proportional to the variation in $S_{\rm gen}$ at all orders in $1/N$, as long as one uses the quantum extremal surface and assumes mild conditions on derivatives of the bulk entropy. The key will be to leverage the fact that $S_{\rm gen}$ is extremized on the QES. Thus, its variation will come from pure boundary terms. At leading order in $z$, we will identify these boundary terms with the vector $V_{\rm QES}$.

We start by varying the generalized entropy with respect to a boundary deformation
\begin{equation}\label{eqn:varySgen}
\delta S_{\rm gen} = \int_{\rm QES} \frac{\delta S_{gen}}{\delta \bar{X}^{i}}\delta \bar{X}^{i} dz d^{d-2}y - \int_{z=\epsilon} \left(\frac{\partial S_{\rm gen}}{\partial (\partial_{z}\bar{X}^{i})} +...\right) \delta \bar{X}^{i}d^{d-2}y
\end{equation}
where the boundary term comes from integrating by parts when deriving the Euler-Lagrange equations for the functional $S_{\rm gen}[\bar{X}]$. The ellipsis denotes terms involving derivatives of $S_{\rm gen}$ with respect to higher derivatives of the embedding functions ($\partial S_{\rm gen}/\partial (\partial^2 X),\ldots$)
These boundary terms will include two types of terms: one involving derivatives of the surface area and one involving derivatives of the bulk entropy. 

The first area term was already calculated in \cite{Koeller:2015qmn}. There it was found that 
\begin{equation}\label{eqn:QNEC3.10}
\frac{\partial A}{\partial(\partial_z\bar{X}^i)} = -\frac{1}{z^{d-1}}\int d^{d-2}y \sqrt{\bar{h}} \frac{g_{ij}\partial_z \bar{X}^i}{\sqrt{1+g_{lm}\partial_z \bar{X}^l\partial_z \bar{X}^m}} \delta \bar{X}^j \vert_{z=\epsilon}
\end{equation}

One can use (\ref{eqn:XQES}) to expand this equation in powers $\epsilon$, and then contract with the null vector $k$ on the boundary in order to isolate the variation with respect to null deformations. For boundary surfaces which are locally stationary at some point $y$, one finds that all terms lower order than $z^d$ vanish at $y$. In fact, it was shown in \cite{Koeller:2015qmn} that the right hand side of (\ref{eqn:QNEC3.10}), after contracting with $k^i$, is just $k^i V_i$ at first non-vanishing order. Finally, we assume that that all such derivatives of the bulk entropy in (\ref{eqn:varySgen}) vanish as $z \to 0$. This is similar to the reasonable assumption that entropy variations vanish at infinity, which should be true in a state with finite bulk entropy. It would be interesting to classify the pathologies of states which violate this assumption. Thus, the final result is that
\begin{equation}\label{eqn:finaleq}
k^i V^{\rm QES}_i = -\frac{1}{\sqrt{h}} k^i\frac{\delta S_{\rm gen}}{\delta X^i}.
\end{equation}
 
The quantum extremal surface prescription says that the boundary field theory entropy is equal to the generalized entropy of the QES \cite{Engelhardt:2014gca}. Setting $S_{\rm gen} = S_{\rm bdry}$ in (\ref{eqn:finaleq}) and combining that with \eqref{eqn:preQNEC} shows that the condition $\beta\geq 0$ is equivalent to the QNEC. Since EWN guarantees that $\beta \geq 0$, the proof is complete.

We briefly comment about the assumptions used to derive (\ref{eqn:finaleq}). The bulk entropy should - for generic states - not depend on the precise form of the region near the boundary. The intuition is clear in the thermodynamic limit where bulk entropy is extensive. As long as we assume strong enough fall-off conditions on bulk matter, the change in the entropy will have to vanish as $z\to 0$.

Note here the importance of using the quantum extremal surface. Had we naively continued to use the extremal area prescription, but still assumed $S_A = S_{\rm bulk}(e(A))+\frac{A}{4G\hbar}$, we would have discovered a correction to the boundary QNEC from the bulk entropy. The variation of the bulk extremal surface area would be given by a pure boundary term, but the QNEC would take the erroneous form
\begin{equation}\label{eqn:wrongQNEC}
T_{kk} \geq \frac{1}{2\pi \sqrt{h}}\left(S_A^{\prime \prime} - S_{bulk}^{\prime \prime}(e(A)) \right).
\end{equation}
In other words, if one wants to preserve the logical connections put forth in Figure \ref{fig:CHART} while accounting for $1/N$ corrections, the use of quantum extremal surfaces is necessary.

We discuss the effects of higher derivative terms in the gravitational action coming from loop corrections at the end of this section.


\subsubsection*{$C \subseteq E$ implies the QHANEC}

We now examine the boundary implication of $\mathcal{C} \subseteq \mathcal{E}$. As before, this proof will hold to all orders in $G\hbar$, again assuming proper fall-off conditions on derivatives of the bulk entropy.

The basic idea will be to realize that general states in AdS/CFT can be treated as perturbations to the vacuum in the limit of small $z$. Again, we will consider the general case where the boundary field theory includes relevant deformations. Then, near the boundary, the metric can be written
\begin{equation}\label{eqn:metricperturbation}
ds^2 = \frac{1}{z^2}\left(dz^2 + \left[f(z)\eta_{ij} + \frac{16\pi G_N}{d}z^dt_{ij}\right]dx^idx^j+o(z^d)\right),
\end{equation}
where $f(z)$ encodes the effects of the relevant deformations. In this proof we take the viewpoint that the order $z^d$ piece of this expansion is a perturbation on top of the vacuum. In other words 
\begin{equation}\label{eqn:pertg}
g_{ab} = g^{vac}_{ab} + \delta g_{ab}.
\end{equation}
Of course, this statement is highly coordinate dependent. In the following calculations, we treat the metric as a field on top of fixed coordinates. We will have to verify the gauge-independence of the final result, and do so below.

For this proof we are interested in regions $A$ of the boundary such that $\partial A$ is a cut of a null plane. In null coordinates, that would look like $\partial A = \lbrace (u= U_0(y),v=0)\rbrace$. These regions are special because in the vacuum state $e(A)$ lies on the past causal horizon generated by bulk geodesics coming from $(u=\infty, v=0)$. This can be shown using Lorentz symmetry as follows:

An arbitrary cut of a null plane can be deformed back to a flat cut by action with an infinite boost (since boosts act by rescalings of $u$ and $v$). Such a transformation preserves the vacuum, and so the bulk geometry possesses an associated Killing vector. The past causal horizon from $(u=\infty, v=0)$ is a Killing horizon for this boost, and by symmetry the quantum extremal surface associated to the flat cut will be the bifurcation surface of the Killing horizon. Had $e(A)$ for the arbitrary cut left the horizon, then it would have been taken off to infinity by the boost and not ended up on the bifurcation surface.\footnote{It is also worth noting that EWN together with $\mathcal{C} \subseteq \mathcal{E}$ can also be used to construct an argument. Suppose we start with a flat cut of a null plane, for which $e(A)$ is also a flat cut of a null plane in the vacuum (the bifurcation surface for the boost Killing horizon). We then deform this cut on the boundary to an arbitrary cut of the null plane in its future. In the bulk, EWN states that $e(A)$ would have to move in a space-like or null fashion, but if it moves in a space-like way, then $\mathcal{C} \subseteq \mathcal{E}$ is violated.}

\begin{figure}
	\centering
	\includegraphics[width= 8cm]{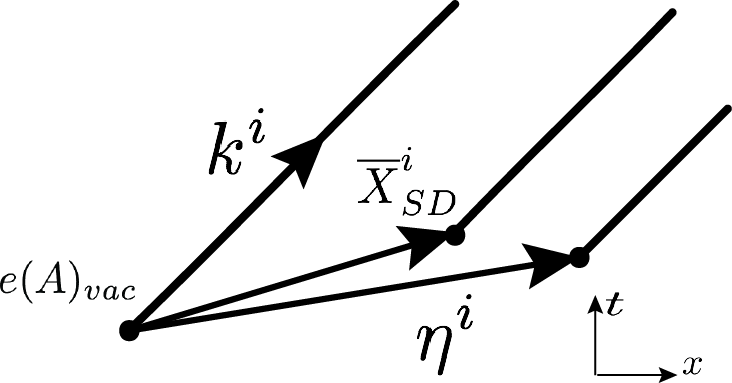}
	\caption{This picture shows the various vectors defined in the proof. It depicts a cross-section of the extremal surface at constant $z$. $e(A)_{vac}$ denotes the extremal surface in the vacuum. For flat cuts of a null plane on the boundary, they agree. For wiggly cuts, they will differ by some multiple of $k^i$.}
    \label{WedgeBCC_fig}
\end{figure}

We can construct an orthogonal null coordinate system around $e(A)$ in the vacuum. We denote the null orthogonal vectors by $k$ and $\ell$ where $k^z = 0 = \ell^z$ and $k^x = k^t = 1$ so that $k\cdot \ell = 1$. Then the statement of $\mathcal{C} \subseteq \mathcal{E}$ becomes\footnote{The issue of gauge invariance for this proof should not be overlooked. On their own, each term in (\ref{eqn:WEDGE}) is not gauge invariant under a general diffeomorphism. The sum of the two, on the other hand, does not transform under coordinate change:
$$
g_{\mu \nu} \to g_{\mu \nu} +\nabla_{(\mu} \xi_{\nu)}
$$
Plugging this into the formula for $k\cdot \eta$ shows that $\delta (k\cdot \eta) = -(k\cdot \xi)$, which is precisely the same as the change in position of the extremal surface $\delta (k\cdot \bar{X}_{\rm SD}) = -(k\cdot \xi)$.}
\begin{equation}\label{eqn:WEDGE}
k\cdot(\eta - \bar{X}_{\rm SD}) \geq 0
\end{equation}
Here we use $\eta, \bar{X}_{\rm SD}$ to denote the perturbation of the causal horizon and quantum extremal surface from their vacuum position, respectively. The notation of $\bar{X}_{\rm SD}$ is used to denote the state-dependent piece of the embedding functions for the extremal surface. Over-bars will denote bulk embedding functions of $e(A)$ surface and $X^a$ will denote boundary coordinates. The set up is illustrated in Figure \ref{WedgeBCC_fig}.

Just as in the previous section, for a locally stationary surface (such as a cut of a null plane), one can write the embedding coordinates of $e(A)$, $\bar{X}$, as an expansion in $z$ \cite{Koeller:2015qmn}:
\begin{equation}\label{eqn:XSD}
\bar{X}^i(y^a,z) = X^i(y^a) + \frac{1}{2(d-2)}z^2 K^i(y^a) +... + \frac{z^d}{d}(V^i + W^i(y^a)\log z) + o(z^d)
\end{equation}
where $V^i$ is some local ``velocity" function that denotes the rate at which the entangling surface diverges from its boundary position and represents the leading term in the state-dependent part of the embedding functions. The state-independent terms of lower order in $z$ are all proportional to $k^i$. In vacuum, we also have $V^i \propto k^i$, and so for non-vacuum states $k \cdot \bar{X}_{\rm SD} = \frac{1}{d}V\cdot k z^d + o(z^d)$.

Equation (\ref{eqn:finaleq}) tells us that $\bar{X}_{SD}$ is proportional to boundary variations of the CFT entropy. Thus, equation (\ref{eqn:XSD}) together with (\ref{eqn:finaleq}) tells us the simple result that 
\begin{equation}\label{eqn:XSD2}
k\cdot \bar{X}_{\rm SD} = -\frac{4G_N}{d\sqrt{h}}S^{\prime}_A z^{d-2}.
\end{equation}

Now we explore the $\eta$ deformation, where $\eta$ is the vector denoting the shift in the position of the causal horizon. This discussion follows much of the formalism found in \cite{Engelhardt:2016aoo}. At a specific value of $(z,y)$, the null generator of the causal surface, $k^\prime$, is related to the vacuum vector $k$ by
\begin{equation}
k^{\prime} = k + \delta k = k + k^a \nabla _a \eta
\end{equation}
In the perturbed metric, $k^{\prime}$ must be null to leading order in $\eta = \mathcal{O}(z^d)$. Imposing this condition we find that
\begin{equation}\label{eqn:deltak}
k^b \nabla_b (\eta \cdot k) = -\frac{1}{2} \delta g_{ab}k^a k^b
\end{equation}
Here $\delta g_{ab}$ is simply the differece between the excited state metric and the vacuum metric, which can be treated as a perturbation since we are in the near-boudnary limit. This equation can be integrated back along the original null geodesic, with the boundary condition imposed that $\eta(\infty) = 0$. Thus, we find the simple relation
\begin{equation}\label{eqn:integratedeta}
(k\cdot\eta)(\lambda) = \frac{1}{2}\int_{\lambda}^{\infty} \delta g_{kk} \,d\tilde\lambda.
\end{equation}

The holographic dictionary tells us how to relate $\delta g_{kk}$ to boundary quantities. Namely, to leading order in $z$, the expression above can be recast in terms of the CFT stress tensor
\begin{equation}
k\cdot \eta = \frac{1}{2}\int_{\lambda}^{\infty} \frac{16\pi G_N}{d} z^{d-2} T_{kk} \,d\tilde\lambda.
\end{equation}
Plugging all of this back in to (\ref{eqn:WEDGE}), we finally arrive at the basic inequality
\begin{equation}\label{eqn:MRE}
\int_{\lambda}^{\infty} T_{kk} \,d\tilde\lambda + \frac{\hbar}{2\pi\sqrt{h}} S^{\prime}_A \geq 0.
\end{equation}
Note that all the factors of $G_N$ have dropped out and we have obtained a purely field-theoretic QHANEC.

\paragraph{Loop corrections}
Here we will briefly comment on how bulk loop corrections affect the argument. Quantum effects do not just require that we add \(S_{\rm out}\) to \(A\); higher derivative terms suppressed by the Planck-scale will be generated in the gravitational action which will modify the gravitational entropy functional. With Planck-scale suppressed higher derivative corrections, derivatives of the boundary entropy of a region have the form
\begin{align}
	S'  = \frac{A'}{4G\hbar} + Q' + S'_{\rm out}
\end{align}
where \(Q'\) are the corrections which start at \(O((G\hbar)^{0})\). The key point is that \(Q'\) is always one order behind \(A'\) in the \(G\hbar\) perturbation theory. As \(G\hbar \to 0\), \(Q'\) can only possibly be relevant in situations where \(A' = 0\) at \(O((G\hbar)^{0})\). In this case, \(V^i \sim k^i\), and the bulk quantum extremal surface in the vacuum state is a cut of a bulk Killing horizon. But then \(Q'\) must be at least \(O(G\hbar)\), since \(Q' = 0\) on a Killing horizon for any higher derivative theory. Thus we find Eq.~\eqref{eqn:finaleq} is unchanged at the leading nontrivial order in \(G\hbar\).

Higher derivative terms in the bulk action will also modify the definition of the boundary stress tensor. The appearance of the stress tensor in the QNEC and QHANEC proofs comes from the fact that it appears at \(O(z^d)\) in the near-boundary expansion of the bulk metric \cite{Koeller:2015qmn}. Higher derivative terms will modify the coefficient of \(T_{ij}\) in this expansion, and therefore in the QNEC and QHANEC. (They will not affect the structure of lower-order terms in the asymptotic metric expansion because there aren't any tensors of appropriate weight besides the flat metric \(\eta_{ij}\) \cite{Koeller:2015qmn}). But the new coefficient will differ from the one in Einstein gravity by the addition of terms containing the higher derivative couplings, which are \(1/N\)-suppressed relative to the Einstein gravity term, and will thus only contribute to the sub-leading parts of the QNEC and QHANEC. Thus the validity of the inequalities at small \(G\hbar\) is unaffected.

\subsubsection*{Boundary Causality Condition implies the AANEC}
The proof of this statement was first described in \cite{Kelly:2014mra}. We direct interested readers to that paper for more detail. Here we will sketch the proof and note some similarities to the previous two subsections.

As discussed above, the BCC states that no bulk null curve can connect boundary points that are not connected by a boundary causal curve. In the same way that we took a boundary limit of $\mathcal{C} \subseteq \mathcal{E}$ to prove the QHANEC, the strategy here is to look at nearly null time-like curves that hug the boundary. These curves will come asymptotically close to beating the boundary null geodesic and so in some sense derive the most stringent condition on the geometry.

Expanding the near boundary metric in powers of $z$, we use holographic renormalization to identify pieces of the metric as the stress tensor
\begin{equation}\label{eqn:holorenorm}
g_{\mu\nu}dx^{\mu}dx^{\nu} = \frac{dz^2+\eta_{ij}dx^{i}dx^{j}+ z^d \gamma_{i j}(z,x^i)dx^{i}dx^{j}}{z^2}
\end{equation}
where $\gamma_{ij}(0,x^i) = \frac{16 \pi G_N}{d}\langle T_{ij} \rangle$. Using null coordinates on the boundary, we can parameterize the example bulk curve by $u \mapsto (u, V(u), Z(u), y^i = 0)$. One constructs a nearly null, time-like curve that starts and ends on the boundary and imposes time delay. If $Z(-L) = Z(L)=0$, then the BCC enforces that $V(L)-V(-L) \geq 0$. For the curve used in \cite{Kelly:2014mra}, the $L\to \infty$ limit turns this inequality directly into the boundary AANEC.

\section{Discussion}\label{sec:discussion}

We have identified two constraints on the bulk geometry, entanglement wedge nesting (EWN) and the $\mathcal{C} \subseteq \mathcal{E}$, coming directly from the consistency of subregion duality and entanglement wedge reconstruction. The former implies the latter, and the latter implies the boundary causality condition (BCC). Additionally, EWN can be understood as a consequence of the quantum focussing conjecture, and $\mathcal{C} \subseteq \mathcal{E}$ follows from the generalized second law. Both statements in turn have implications for the strongly-coupled large-\(N\) theory living on the boundary: the QNEC and QHANEC, respectively. In this section, we list possible generalizations and extensions to this work.

\paragraph{Unsuppressed higher derivative corrections}

There is no guarantee that higher derivative terms with un-suppressed coefficients are consistent with our conclusions. In fact, in  \cite{Camanho:2014apa} it was observed that Gauss-Bonnett gravity in AdS with an intermediate-scale coupling violates the BCC, and this fact was used to place constraints on the theory. We have seen that the geometrical conditions EWN and $\mathcal{C} \subseteq \mathcal{E}$ are fundamental to the proper functioning of the bulk/boundary duality. If it turns out that a higher derivative theory invalidates some of our conclusions, it seems more likely that this would be point to a particular pathology of that theory rather than an inconsistency of our results. It would be interesting if EWN and $\mathcal{C} \subseteq \mathcal{E}$ could be used to place constraints on higher derivative couplings, in the spirit of \cite{Camanho:2014apa}. We leave this interesting possibility to future work.

\paragraph{A further constraint from subregion duality}
Entanglement wedge reconstruction implies an additional property that we have not mentioned. Given two boundary regions $A$ and $B$ that are spacelike separated, $\mathcal{E}(A)$ is spacelike separated from $\mathcal{E}(B)$. This property is actually equivalent to EWN for pure states, but is a separate statement for mixed states. In the latter case, it would be interesting to explore the logical relationships of this property to the constraints in Fig.~\ref{fig:CHART}.

\paragraph{Beyond AdS}
In this paper we have only discussed holography in asymptotically AdS spacetimes. While the QFC, QNEC, and GSL make no reference to asymptotically AdS spacetimes, EWN and $\mathcal{C}\subseteq \mathcal{E}$ currently only have meaning in this context. One could imagine however that a holographic correspondence with subregion duality makes sense in more general spacetimes --- perhaps formulated in terms of a ``theory" living on a holographic screen \cite{Bousso:1999cb,Bousso:2015mqa,Bousso:2015qqa}. In this case, we expect analogues of EWN and $\mathcal{C}\subseteq \mathcal{E}$. For some initial steps in this direction, see \cite{Sanches:2016aa}.

\paragraph{Quantum generalizations of other bulk facts from generalized entropy}
A key lesson of this paper is that classical results in AdS/CFT relying on the null energy condition (NEC) can often be made semiclassical by appealing to powerful properties of the generalized entropy: the quantum focussing conjecture and the generalized second law. We expect this to be more general than the semiclassical proofs of EWN and $\mathcal{C} \subseteq \mathcal{E}$ presented here. Indeed, Wall has shown that the generalized second law implies semiclassical generalizations of many celebrated results in classical general relativity, including the singularity theorem \cite{C:2013uza}. It would be illuminating to see how general this pattern is, both in and out of AdS/CFT. As an example, it is known that strong subadditivity of holographic entanglement entropy can be violated in spacetimes which don't obey the NEC \cite{Callan:2012ip}. It seems likely that the QFC can be used to derive strong subadditivity in cases where the NEC is violated due to quantum effects in the bulk.

\paragraph{Gravitational inequalities from field theory inequalities}
We have seen that the bulk QFC and GSL, which are semi-classical quantum gravity inequalities, imply their non-gravitational limits on the boundary, the QNEC and QHANEC. But we can regard the bulk as an effective field theory of perturbative quantum gravity coupled to matter, and can consider the QNEC and QHANEC for the bulk matter sector. At least when including linearized backreaction of fields quantized on top of a Killing horizon, the QHANEC implies the GSL \cite{Wall:2011hj}, and the QNEC implies the QFC \cite{Bousso:2015mna}. In some sense, this ``completes'' the logical relations of Fig.~\ref{fig:CHART}.

\paragraph{Support for the quantum extremal surfaces conjecture}
The logical structure uncovered in this paper relies heavily on the conjecture that the entanglement wedge should be defined in terms of the surface which extremizes the generalized entropy to one side \cite{Engelhardt:2014gca} (as opposed to the area). Perhaps similar arguments could be used to prove this conjecture, or at least find an explicit example where extremizing the area is inconsistent with subregion duality, as in \cite{Gao:2016bin}.

\paragraph{Connections to Recent Proofs of the AANEC}
Recent proofs of the AANEC have illuminated the origin of this statement within field theory \cite{Faulkner:2016mzt, Hartman:2016lgu}. In one proof, the engine of the inequality came from microcausality and reflection positivity. In the other, the proof relied on montonoicity of relative entropy for half spaces. A natural next question would be how these two proofs are related, if at all. Our paper seems to offer at least a partial answer for holographic CFTs. Both the monotonicity of relative entropy and microcausality - in our case the QHANEC and BCC, respectively - are implied by the same thing in the bulk: $\mathcal{C} \subseteq \mathcal{E}$. In \ref{sec:glossary}, we gave a motivation for this geometric constraint from subregion duality. It would be interesting to see how the statement of $\mathcal{C} \subseteq \mathcal{E}$ in a purely field theoretic language is connected to both the QHANEC and causality.


\section*{Acknowledgements}
It is a pleasure to thank Raphael~Bousso, Venkatesa~Chandrasekaran, Zach~Fisher, Illan~Halpern, Arvin~Shahbazi~Moghaddam, and Aron~Wall for discussions. 
Our work is supported in part by the Berkeley Center for Theoretical Physics, by the National Science Foundation (award numbers 1214644, 1316783, and 1521446), by fqxi grant RFP3-1323, and by the US Department of Energy under Contract DE-AC02-05CH11231.

\bibliographystyle{utcaps}
\bibliography{all}

\end{document}